\documentclass[12pt]{iopart}
\newcommand{\bd}[1]{ \mbox{\boldmath $#1$}  }
\newcommand{\nn}{\nonumber}
\newcommand{\beq}{\begin{equation}}
\newcommand{\eeq}{\end{equation}}
\newcommand{\beqa}{\begin{eqnarray}}
\newcommand{\eeqa}{\end{eqnarray}}
\newcommand{\ba}{\begin{array}}
\newcommand{\ea}{\end{array}}
\newcommand{\alp}{\alpha}
\newcommand{\lam}{\lambda}

\newcommand{\NPA}{Nucl. Phys. {\bf A}}

\newcommand{\PRC}{Phys. Rev. C}

\usepackage{iopams}  
\usepackage{epsf}
\usepackage{color}
\begin{document}
 
\title[$\alpha$ Decay in Ultra-Intense Laser Fields]
{$\alpha$ Decay in Ultra-Intense Laser Fields}

\author{\c S. Mi\c sicu \footnote[8]{To
whom correspondence should be addressed (misicu@theory.nipne.ro)} and M. Rizea  \footnote[3]{(rizea@theory.nipne.ro)}}

\address{Department for Theoretical Physics, National Institute 
for Nuclear Physics and Engineering - HH, Bucharest-M\u agurele, POB MG-6,
Romania}

\begin{abstract}
We investigate the $\alpha$-decay of a spherical nucleus under the influence of an ultra-intense laser field
for the case when the radius vector joining the center-of-masses of the $\alpha$-particle and the daughter is
aligned with the direction of the external field.     
The time-independent part of the $\alpha$-daughter interaction is taken from elastic scattering compilations
whereas the time-varying part describes the interaction between the decaying system with the laser field. 
The time-dependent Schr\"odinger equation is solved numerically by appealing to a modified scheme of the Crank-Nicolson type 
where an additional first-order time derivative appears compared to the field-free case. 
The tunneling probability of the $\alpha$-cluster, and derived quantities (decay rate, total flux) is determined for various 
laser intensities and frequencies for either continous waves or few-cycle pulses of envelope function $F(t)=1$. 
We show that in the latter case pulse sequences containing an odd number of half-cycles determine an enhancement of the 
tunneling probability compared to the field-free case and the continuous wave case.
The present study is carried out taking as example the alpha decaying nucleus $^{106}$Te.
\end{abstract}
\maketitle

\section{Introduction}

Nowadays laser fields with high intensities and a wide range of frequencies are available in the form of
short pulses. 
The advent of chirped pulse amplification (CPA) techniques along with the development of high-fluence laser materials
resulted in the production of electromagnetic field intensities in excess of 10$^{18}$ W/cm$^{2}$ \cite{mour06}. 
Such lasers fields are already strong enough to compete with Coulomb forces inside atoms and 
consequently they presently allow the control of processes at this level: multiphoton and above-threshold 
ionization, harmonic generation and attosecond pulses, laser cooling and laser trapping
of atoms, laser-assisted electron-atom collisions \cite{brab08}.

However the highest attained laser intensities as well the lowest photon wavelengths photon energies are still 
too far from exploring strongly bound nuclear matter. In order to overcome the distance on the energy scale between the
atomic and the nuclear levels, a new type of large-scale laser facility is under construction near Bucharest \cite{eli09}.
The ambitious task of this european joint project is to produce pulses of wavelength $\lam$=800 nm (Ti:saphire laser), 
duration $\tau_p$=10 fs, energy release of a few kJ, and to focus each shot into a 1$\times$1 $\mu$m$^2$ spot.  
Under such circumstances, the Extreme Light Infrastructure (ELI) facility is claimed to be capable 
to yield a peak power of $P_{\rm max}\approx$ 200 PW and an intensity as high as $I_0=10^{25}$W/cm$^{2}$.

In spite of the great expectations unleashed by the entrance into operation in the foreseeable future of this extremely 
powerfull machine one should not overlook that we are still confronted with the problem that on one hand many low-energy 
nuclear excitations involve energies in the range 10$^{-3}$-10 MeV whereas laser photon energies, such as those produced 
with the Ti:saphire ($\lam$=800 nm), are barely exceeding 1 eV. 
On the other hand the X-ray free-electron lasers (XFEL) are able to produce photon energies above 10 keV 
(see more about this subject in \cite{schmus08,xfel11}).
For example the Linac Coherent Light Source (LCLS) \cite{ishik12} 
succeeds the generation of sub-angstrom laser light by combining a short-period 
undulator with an 8 GeV electron beam, the maximum power barely exceeding 10 GW with a pulse duration of 10 fs. 
The shortest wavelength attained 0.634{\AA} corresponds to a photon energy of 19.6 keV. 
Another XFEL project is planned to be the SwissFEL facility where pulses of X-rays in the wavelengths range 1-70 \AA~ are 
produced with a pulse duration in the range 5-20 fs, a pulse energy of the order of a mJ, each shot being focused 
into a 100$\times$100 nm$^2$ spot. Consequently a peak power below 1 TW and an intensity not higher than 
$10^{22}$W/cm$^{2}$ are expected \cite{swiss11}. It would be ideal to have both the ultra-high intensities produced by ELI and
the low wavelengths of XFEL produced with the same machine. 
  
On the theoretical side there is an increasing focus on direct laser-nucleus reactions (see \cite{piazza12} and references therein).
A very recent work discussed in a semi-classical framework the laser-induced recollisions suffered by an $\alpha$ particle spontaneously 
emitted in the spontaneous decay of a heavy nucleus \cite{casta11}. Consequently the variation of the $\alpha$ particle kinetic 
energy results from the classical dynamics in a time-varying potential and not by quantum jumps from the ground-(metastable) states
to higher-lying states in the continuum. The application was made for laser intensities of  $10^{22}-10^{23}$ W/cm$^{2}$ 
and photon frequencies produced by a Ti:saphire laser. It was concluded that relative modifications of
 $\alpha$-decay half-lives, though small, are possible under the laser influence.

Attempts to modify $\alpha$ decay rates were made already soon after the discovery of radioactivity. 
In one of these experiments, Rutherford attempted a sudden change of temperature and pressure, up to
2500 $^0$C and pressures of 100 MPa for a short period of time, and no effect was observed in the 
half-lives (as cited in \cite{schw06}). This negative result could be explained in the Gamow picture of quantum 
mechanical tunneling : the distortions or even removal of the electron cloud surrounding the mother nucleus  caused 
by temperature, pressure, magnetic or gravitational fields, are producing a negligible effect on the Coulomb barrier,
such that changes in the decay rate of the order of $\delta\lambda/\lambda\approx 10^{-7}$ are expected \cite{emery72}.  
Even some years ago it has been suggested that one can speed up $\alp$ decay of transuranic nuclear waste
by embedding in it metals at low temperature \cite{kett06}, a proposal that was soon refuted based on  the 
ground that medium modifications are affecting the alpha-daughter potential mainly at radii larger than the outer turning point 
\cite{zinn07}.  

In this work we investigate the effect of a linearly polarized, ultra-intense laser field on the tunneling probability 
of an $\alpha$ particle emitted from a heavy spherical nucleus. Consequently we adopt a non-perturbative formalism
adequate for strong fields, i.e. we perform a direct numerical integration of the time-dependent Schr\"odinger equation (TDSE)
describing the relative alpha cluster - daughter nucleus motion under the effect of a continuous wave (cw) or a sequence of ultrashort
pulses containing a few even or odd number of half-cycles of duration no longer than 1 as. 
The latter issue is related to the opportunity, available after the turn of the millenium,
to access parameter ranges in high-field physics by using laser pulses comprising a small number of cycles \cite{brab00}. 
Phase-controlled light pulses are expected to allow control of high-intensity 
light-matter interactions on a sub-cycle time scale.
Our primary goal is to determine how the interaction of radiation with this 
metastable dinuclear system could speed-up or hinder its decay.

{As a case study we consider the $\alp$-decay of the nucleus $^{106}$Te. 
Along with other $\alp$-emitters close above the double-magic nucleus $^{100}$Sn, this
nucleus has been a subject of experimental investigations in recent times \cite{jan05,sew06}.
Measured $\alp$-decay energies of trans-tin emitters increase along the decay chains as a result
of particularly strong binding of nuclei with $Z\approx N$ in this mass region, a conclusion
confirmed by WKB penetrability calculations \cite{xu06,mo07}.}

\section{Dinuclear System in a Laser Field}

The aim of this section is to derive the solution of the initial value problem for the 
Schr\"odinger equation describing two charged nuclei, initially found in a quasi-bound state, 
and perturbed by a time-varying external field  
\beq
i\hbar \frac{\partial\psi({\bi r}_1,{\bi r}_2,t)}{\partial t}=H(t)\psi({\bi r}_1,{\bi r}_2,t)
\label{tdse1}
\eeq
The Hamiltonian of the cluster ($\alpha$) - daughter ($d$) nucleus system with masses   $m_1=m_\alp$ and $m_2=M_d=M-m_1$,  
charges $Z_1=2$, $Z_2=Z-2$, with positions in the laboratory frame specified  by $\bi{r}_1$ and $\bi{r}_2$  and momenta 
$\bi{p}_1$ and $\bi{p}_2$, coupled to the electromagnetic field, described by the transverse vector potential ${\bi A}$, 
reads in the Coulomb gauge as \cite{cohen97} :
\beq
H(t)=\frac{\bi{p}_{1}^2}{2m_1}+\frac{\bi{p}_2^2}{2m_2}+V(|\bi{r}_1-\bi{r}_2|)+H_{\rm int}(t)
\eeq 
The time-independent part of the above Hamiltonian describes the $\alp$-daughter system interacting via 
Coulomb+nuclear potential $V$ in the absence of the laser field. The Hamiltonian describing the 
interaction of the two nuclear charges with the laser field is given by
\beq
H_{\rm int}(t)=-\sum_{i=1,2}\frac{Z_{i}e}{m_{i}}{\bi{p}_{i}\cdot\bi{A}(\bi{r}_{i},t)}
+\sum_{i=1,2}\frac{(Z_{i} e)^2}{2 m_{i}}{\bi A}^2(\bi{r}_{i},t)
\label{inter1}
\eeq
The transformation to the center-of-mass coordinates $({\bi R},{\bi r})$ and momenta  $({\bi P},{\bi p})$
is defined by the set of relations :
\beq
{\bi r}_1={\bi R}+\frac{m_2}{m_1+m_2}{\bi r},~~~{\bi r}_2={\bi R}-\frac{m_1}{m_1+m_2}{\bi r}
\label{pozcmrel}
\eeq
\beq
{\bi p}_1={\bi p}+\frac{m_1}{m_1+m_2}{\bi P},~~~{\bi p}_2=-{\bi p}+\frac{m_2}{m_1+m_2}{\bi P}
\label{momcmrel}
\eeq
In this particular case, the two charges are assumed to form a quasi-bound state whose dimensions, 
$R_{N}\sim r_0 (A_1^{1/3}+A_2^{1/3})$ are small compared to the radiation wavelength $\lam=2\pi c/\omega$.
Proceeding in full analogy to the atomic case, we perform the calculations in the long wavelength approximation, i.e.
we assume that $k R_N\ll 1$ and that the vector potential is spatially homogenous. Thus ${\bi A}({\bi r}_1)
\approx {\bi A}({\bi r}_2)\approx{\bi A}({0})$  ({\em dipole approximation}) and the interaction Hamiltonian 
is rewritten in the C.M. coordinates as
\beq
H_{\rm int}(t)=-\frac{e}{m}\left [ \frac{Z}{A}{\bi P} 
+ \left ( \frac{Z_1}{A_1} - \frac{Z_2}{A_2}\right ){\bi p}\right ]\cdot{\bi A}
+\frac{e^2}{2m}\left ( \frac{Z^2_1}{A_1} +  \frac{Z^2_2}{A_2}  \right ){\bi A}^2
\eeq
where $m$ is the nucleon mass. 

Substituting the transformations (\ref{pozcmrel})-(\ref{momcmrel}) into (\ref{tdse1}) we obtain the TDSE in the dipole approximation 
in the form 
\beq
i\hbar \frac{\partial\psi({\bi r},{\bi R},t)}{\partial t}=\left [ \frac{{\bi p}^2}{2\mu}+\frac{{\bi P}^2}{2 M}
+V({\bi r}) + H_{\rm int}\right ]\psi({\bi r},{\bi R},t)
\eeq
The above TDSE can be simplified by an appropriate quantum mechanical gauge transformation. Applying a generalized G\"oppert-Mayer 
transformation to the above equation yields a vanishing vector potential ({\em length gauge}) \cite{cohen97}. 
The wave function is transformed according to  
\beq
\psi({\bi r},{\bi R},t)={\hat U}_r {\hat U}_R{\psi^{\rm L}}({\bi r},{\bi R},t)
\eeq  
where 
\beq
{\hat U}_r=\exp\left [ \frac{i}{\hbar} eZ_{\rm eff}{\bi A}\cdot{\bi r} \right ],
~~~{\hat U}_R=\exp\left [ \frac{i}{\hbar} eZ{\bi A}\cdot{\bi R} \right ]
\eeq
Let us introduce the transverse electric field 
\beq
{\bi E}(t)=-\frac{\partial{\bi A}(t)}{\partial t}
\label{ele-avec}
\eeq
The two-body Schr\"odinger equation in the {\it length gauge} acquires then the simple form
\beqa
i\hbar \frac{\partial{\psi^{\rm L}}({\bi r},{\bi R},t)}{\partial t}&=&
\left [ -\frac{\hbar^2}{2\mu}\nabla^2_{r} - eZ_{\rm eff}{\bi r}\cdot{\bi E}(t) +V({\bi r})\right.\nn\\  
&&\left.        -\frac{\hbar^2}{2 M} \nabla^2_{R} - eZ{\bi R}\cdot{\bi E}(t) \right ]{\psi^{\rm L}}({\bi r},{\bi R},t)
\label{eqlegaug}
\eeqa
where 
$$Z_{\rm eff}=\frac{Z_1A_2-Z_2A_1}{A_1+A_2},~~~~\mu=\frac{m_1m_2}{m_1+m_2}$$
Thus in the {\it length gauge} the interaction Hamiltonian couples the C.M. $\bi{R}$ and the relative ${\bi r}$ coordinates
to the electric field ${\bi E}(t)$.
This equation can be separated and in order to accomplish this task we introduce the factorization of the wave-function 
\beq
{\psi^{\rm L}}({\bi r},{\bi R},t)=\phi^{\rm L}({\bi r},t)\chi^{\rm L}({\bi R},t).
\eeq 
The two separated time-dependent parabolic equations are
\beq
i\hbar \frac{\partial{\chi^{\rm L}}({\bi R},t)}{\partial t}=
\left [-\frac{\hbar^2}{2 M} \nabla^2_{R} - eZ{\bi R}\cdot{\bi E}(t) \right ]{\chi^{\rm L}}({\bi R},t)
\eeq
for the center-of-mass motion and
\beq
i\hbar \frac{\partial{\phi^{\rm L}}({\bi r},t)}{\partial t}=
\left [ -\frac{\hbar^2}{2\mu}\nabla^2_{r} +V({\bi r})-eZ_{\rm eff}{\bi r}\cdot{\bi E}(t)\right ]{\phi^{\rm L}}({\bi r},t)
\label{tdserel2}
\eeq
for the relative coordinate, describing   the relative $\alpha$-cluster - daughter motion in an electromagnetic field. 
The dynamics of the C.M., i.e. the ``free'' dinuclear system in the presence of a radiation field $\bd{A}(t)$, is described by 
Volkov states and is richly exposed elsewhere (see for example \cite{mittle93}).
Since in the present approach the decay trajectory is assumed to be colinear with the direction of the laser field, the C.M. 
wave-function $\chi$ is fully decoupled from the $r$-dependent decay wave-function $\phi$. Translations or rotations as a whole of this system 
are therefore not the subject of subsequent considerations. In the case of a field with arbitrary polarization this 
approximation is no longer satisfactory.

A furher simplification of TDSE is achieved by removing the term in ${\bi A}^2$.
This can be done by means of a unitary operator which singles out a time-dependent phase factor from the wave function 
\beq
{\hat U}_{\rm V}=\exp\left [ \frac{i}{\hbar} \frac{e^2}{2m}
\left (\frac{Z^2_{1}}{A_1}+\frac{Z^2_{2}}{A_2}\right )\int dt'{\bi A}^2(t') \right ]
\eeq     
and the TDSE in the relative-coordinate reads
\beq
i\hbar \frac{\partial{\phi^{\rm V}}({\bi r},t)}{\partial t}=
\left [ -\frac{\hbar^2}{2\mu}\nabla^2_{r} +V({\bi r})
-{\frac{1}{\mu}}eZ_{\rm eff}{\bi p}\cdot{\bi A}(t)\right ]{\phi^{\rm V}}({\bi r},t)
\eeq 
which is known as {\it velocity gauge} since the interaction Hamiltonian couples relative velocity ${\bi p}/\mu$ to 
the vector potential.

Performing a time-dependent unitary transformation on top of the above {\em velocity-gauge} wave function removes the 
${\bi p}\cdot{\bi A}$ term in the interacting part of the Hamiltonian \cite{henneb68}. The relative motion wave-function in the new representation
(Kramers-Henneberger) is related to the old by means of
\beq
\phi^{\rm V}({\bi r},t)=\exp\left [ -\frac{i}{\hbar}{\bd{\alpha}}(t)\cdot{\bi p}\right ]\phi^{\rm K-H}({\bi r},t)
\eeq
where 
\beq
{\bd{\alpha}}(t)=-\frac{eZ_{\rm eff}}{\mu}\int dt'{\bi A}(t') 
\eeq
is the vector corresponding to the displacement of the $\alpha$-cluster from its oscillation centre in the electric field 
${\bd{E}(t)}$ and in atomic physics is known also under the name of {\em quiver} motion.  

The TDSE in the oscillating Kramers-Henneberger frame reads then
\beq
i\hbar \frac{\partial{\phi^{\rm K-H}}({\bi r},t)}{\partial t}=
\left [ -\frac{\hbar^2}{2\mu}\nabla^2_{r} +V({\bi r}+\bd{\alpha}(t))\right ]{\phi^{\rm K-H}}({\bi r},t)
\label{tdsekh}
\eeq
The above form shows that in the K-H  representation the effect of the external field is completely transferred into the argument 
of the static nuclear+Coulomb potential. Eq.(\ref{tdsekh}) provides the dynamics of the $\alpha$-cluster in a moving frame of reference
which follows the quiver motion $\bd{\alpha}(t)$. In the K-H frame of reference the daughter nucleus has instead a quiver motion
$-\bd{\alpha}(t)$ \cite{gavr92}.

\section{$\alpha$ decay in a linear-polarized, monochromatic laser field}

Some of the methods developed at the atomic level prove to be useful 
in the modelling of the laser-nucleus interaction \cite{gross08}. For a laser pulse with a peak of the electric field 
much smaller than the electric fields experienced by an $\alp$-cluster 
on the surface of a heavy nucleus, the theoretical
treatment can be done by resorting to perturbative methods. In the present paper we consider ultra-intense laser fields,
a fact that makes our problem tractable in a non-perturbative framework. Most popular non-perturbative 
methods consist in the direct numerical integration of the Schr\"odinger equation and have the advantage that 
solutions can be obtained for a wide range of laser intensities and frequencies, regardless of the pulse type.

The $\alpha$-decay in a ultra-intense laser field is treated below in a quantum-mechanical approach under the 
provision of two simplifying assumptions. We first assume that the electric field is linearly polarized and its direction
$\bd{E}(t)={\cal E}_0(t)\bi{e}_x$ along the $x$-axis is parallel to the symmetry axis of the dinuclear system, i.e. we 
select a fixed orientation in space of the decaying system and discard from further considerations all possible non-axial 
configurations with respect to the field. We postpone for a future study the two-dimensional approach to this problem and 
adopt in the present paper a one-dimensional geometry. In this approximation the TDSE describing the tunneling process in the 
{\em length-gauge} (\ref{tdserel2}) is recasted as follows
\beq
i\hbar \frac{\partial{\phi}(x,t)}{\partial t}=
\left [ -\frac{\hbar^2}{2\mu}\frac{d^2}{dx^2} +V(x)+V_{\rm int}(x,t))\right ]{\phi}({x},t)
\label{tdse1dim}
\eeq    
In the case discussed in this paper the above equation describes the one-dimensional quantum motion of an $\alpha$-particle inside a static 
potential $V(x)$ subjected during the tunneling process to an additional time-dependent external field, that according to eq.(\ref{eqlegaug})
reads
\beq
V_{\rm int}=- eZ_{\rm eff}{x}{E}(t)
\eeq 
The static $\alp$-daughter $V(r)$ potential comprises a long range Coulomb potential and the nuclear potential in the 
Woods-Saxon form
\beq
V_{\rm nucl}=-\frac{V_0}{1+e^{(|x|-R_n)/a}}
\label{eq_ws}
\eeq
where $V_0$ is the depth, $R_n=r_0\times A_2^{1/3}$ and $a$ the potential diffuseness.   
The electric field of a non-dispersive laser can be represented by a modulated linear-polarized and monochromatic plane wavefunction
(single-mode field) 
\beq
{E}(t)={\cal E}_0 {F}(t)\sin(\omega t+\delta)
\eeq
where ${\cal E}_0$ is the electric field strength, $F(t)$ is the pulse shape function (envelope) and 
$\omega=2\pi/T$ is the carrier angular frequency. In practical situations $F(t)$ is taken to vary between 0 and 1, 
and thus ${\cal E}_0$ is the peak of the pulse. The carrier-envelope phase (CEP) $\delta$ is the phase of the
carrier wave with respect to the maximum $F(t)$. We assume in what follows that a maximum of the laser pulse envelope occurs 
at $t=0$. The envelope phase $\delta$ is meaningless in what concerns the response of the decaying system when the pulses are long.

Let us return now to eq.(\ref{tdse1dim}).
The dipole interaction $V_{\rm int}$ is proportional to $x$, a fact that inherently leads to numerical instabilities 
at the computational boundaries. To remove this term we convert the Hamiltonian using the transformation
\cite{chou12} 
\beq
\phi(x,t)=\exp\left (-\frac{i}{\hbar}{eZ_{\rm eff} A(t)\cdot x}\right ){\widetilde \phi}(x,t)
\eeq 
In the new form, the TDSE reads:
\beq
i\hbar\frac{\partial{\widetilde \phi}}{\partial t} =
\frac{1}{2\mu}\left [ \left ( i\hbar\frac{\partial}{\partial x}
-eZ_{\rm eff}A(t)\right )^2{\widetilde \phi}\right ] +V(x){\widetilde \phi}
\label{tdsenew}
\eeq
The direct numerical integration of the above wave equation is done by means of the Crank-Nicolson(CN) 
method \cite{crank47}. It is deduced by writing the formal solution of the above equation:
\begin{equation}
{\widetilde \phi}(x,t+\Delta t) = e ^{ -{{i \Delta t}\over \hbar} H(x,t)}
{\widetilde \phi}(x,t)
\label{eqcn}
\end{equation}
where $H(x,t)$ is the hamiltonian given by
\begin{equation}
H =  \frac{1}{2\mu}\left ( i\hbar\frac{\partial}{\partial x}
-eZ_{\rm eff}A(t)\right )^2 +V(x)
\end{equation}
Then, expanding $H$ in Taylor series up to the second-order in $t$ and using the Pad\'{e} approximation, 
the exponential in eq.(\ref{eqcn}) is expressed as the ratio of two first order 
polynomials (the Cayley transformation)
\begin{equation}
e ^{-{{i \Delta t}\over \hbar} H} \approx 
{{1 - {{i \Delta t} \over {2 \hbar}}H  - {{i \Delta t^2} \over {4 \hbar}}{\dot H}     } 
\over {1 + {{i \Delta t} \over {2 \hbar}}H }+ {{i \Delta t^2} \over {4 \hbar}}{\dot H} },~~~{\dot H}=\frac{\partial H}{\partial t}
\end{equation}
so that the solution at the moment $t + \Delta t$ 
(denoted by ${\widetilde \phi}^{n+1}$) is obtained from the solution at the 
moment $t$ (denoted by ${\widetilde \phi}^{n}$)  according to the formula
\begin{equation}
\left ( 1 + {{i \Delta t} \over {2 \hbar}} H + {{i \Delta t^2} \over {4 \hbar}}{\dot H} \right ) {\widetilde \phi}^{n+1} = 
\left ( 1 - {{i \Delta t} \over {2 \hbar}} H - {{i \Delta t^2} \over {4 \hbar}}{\dot H} \right ) {\widetilde \phi}^n
\end{equation}
The error in time is proportional with $\Delta t ^3$.
In practice, the derivatives with respect to $x$ appearing in $H$ and ${\dot H}$ are approximated by 
finite differences on spatial mesh points and the solution at each time step
is obtained by solving a linear system.
 
The CN scheme has a number of attractive properties : the initial wave function is required only at the
starting value of time, it is unconditionally stable, it is unitary and conserves the norm \cite{ask78}. 
Moreover, if the derivatives with respect to $x$ are approximated by 
the usual 3-point formulas, at each time step a tridiagonal linear system should be solved,
which can be done fast and accurate up to the machine precision.

Note that when the Hamiltonian does not contain the first derivative with respect to $x$, the spatial accuracy can
be improved either by using Numerov-like formulas (see \cite{moy04,riz10}), or by approximating 
the second derivative with higher order finite difference formulas \cite{dijk07}. In the later case, more
complicated linear systems result (with the number of diagonals larger than 3) during the time propagation.
In the present case, both derivatives (first and second) appear in the Hamiltonian and to increase the spatial
accuracy, finite difference formulas in more points should be used for each derivative, leading to non-tridiagonal systems 
which require longer computational effort at each time step.
The accuracy in time could also be increased by using the Magnus expansion and diagonal Pad\'e approximation \cite{puz99}.
Again, this leads to additional computing time. For example, for a scheme of order $(\Delta t)^5$, a pair of linear systems should
be solved in order to advance one step in time \cite{riz08}.

The physical problem considered in this paper necessitates a carefull treatment of the space and time grids. 
Due to the high laser intensities the $\alp$-cluster can acquire a quiver motion $\alp(t)$
much larger than the initial average size of the dinuclear system and therefore the 
forerunners of the wave function can travel large distances in short times. Consequently the 
spatial grids used in the integration scheme must be correspondingly large and have a 
short step in order to control as good as possible the fine changes in the wavefunction. 
For the sake of accuracy the time discretization requires also a large number of small steps.  
In the numerical calculations we use a spatial mesh size $\Delta x$ of 1/8 fm and time 
step $\Delta t$, depending on the practical conditions imposed by the initial wave function or the 
laser field  parameters. The value $\Delta t=1/8$  in units of 10$^{-22}$s was satisfactory for most
calculations. For testing we used smaller spatial and temporal steps, obtaining the same behavior 
and very close results. The maximum time limit we achieved was $\sim 10^{-15}$s which is roughly 
4 orders of magnitude higher than the limit attained previously by us \cite{mis00}.
This is obvious an improvement in comparison with other TDSE  approaches 
to the $\alpha$-decay problem which can be found in the literature.

When numerically solving TDSE over the real line $(-\infty,\infty)$, the computation has to be confined
to a finite domain, say ${\cal D}=[x_{\rm left},x_{\rm right}]$, by introducing artificial boundary conditions. 
As we shall see below, the intial data is supported on this finite domain. Consequently the exact 
solution of the whole real axis problem restricted to ${\cal D}$, can be approximated by solving the TDSE 
only on ${\cal D}$, together with artificial boundary conditions at $x_{\rm left}$ and $x_{\rm right}$.
If this approximate solution coincides on ${\cal D}$ with the exact solution, the boundary conditions are dubbed
as {\em Transparent Boundary Conditions} (TBC).     
A proper implementation of this algorithm is paramount in order to avoid reflections of the propagated wave-function at 
the grid frontiers and thus causing errors in the calculation of physical quantites.
In this work we use TBC in the form suggested in \cite{hadl91} and before implementing this algorithm 
to our specific problem let us introduce some useful quantites. 
Any solution of the TDSE (\ref{tdsenew}) satisfies the continuity equation for the probability density :
\beq
\frac{\partial \rho}{\partial t}+\frac{d}{dx}{J}=0
\eeq
where $\rho$ is the probability density 
\beq
\rho(x,t)=|\widetilde \phi(x,t)|^2
\eeq
and $J$ is one-dimensional current or probability flux
\beq
J(x,t)=-\frac{i\hbar}{2\mu}\left [ {\widetilde \phi^*(x,t)} \frac{d}{dx}{\widetilde \phi(x,t)} -  
{\widetilde \phi(x,t)} \frac{d}{dx}{\widetilde \phi^*(x,t)}\right ]
+\frac{eZ_{\rm eff}}{\mu}A(t)\rho(x,t)\nn\\
\eeq
as can be checked by direct calculation.
  
The time-dependent tunneling probability measuring the escape likelihood of the $\alpha$-particle
from the nuclear+Coulomb potential well is defined by
\beq
P_{\rm tun}(t)=\left ( \int_{x_b}^{\infty}+\int_{-\infty}^{-x_b}\right )\rho(x,t)dx=1-\int_{-x_b}^{x_b} \rho(x,t)dx
\label{probtun}
\eeq
where $\pm x_b$ are the static barrier positions. 
If we take the derivative with respect to $t$ of the above equation and use the continuity equation we obtain a relation between
the tunneling rate and the flux across the nuclear surface.
\beq
{\dot P}_{\rm tun}(t)=J(x_b,t)-J(-x_b,t)
\eeq
  Mutatis mutandis, if we introduce the total norm inside the numerical grid, i.e. we integrate the probability density
between the left, $x_{\rm left}$, and the right $x_{\rm right}$ boundaries:
\beq
P_{\rm int}(t)=\int_{x_{\rm left}}^{x_{\rm right}}\rho(x,t)dx
\label{pinterior}
\eeq
we obtain the flux across the numerical grid
\beq
{\dot P}_{\rm int}(t)=J({x_{\rm left}},t)-J({x_{\rm right}},t)
\label{balans}
\eeq
Since the treatment of the two boundaries is identical we continue our considerations
by focusing only on the right boundary. Following \cite{hadl91} we make an essential
assumption, i.e. near this boundary ${\widetilde \phi}=A\exp(ik_x\cdot x)$, where $A$ and
$k_x$ are complex constants. With this assumption the flux leaving the right boundary 
$x_{\rm right}$ is :
\begin{equation}
{\it J}(x_{\rm right},t) = \frac{1}{\mu}
\left [{\rm Real}\left (\hbar k_x \right )+eZ_{\rm eff}A(t)\right ]
|{\widetilde \phi}(x_{\rm right},t)|^2
\label{fluxr}
\end{equation}
An important feature of this procedure is that $k_x$ is allowed to change
as the problem progresses, thus eliminating the need for a problem-dependent
adjustable parameter.
If $x_M$ is the boundary of the spatial grid, the above mentioned assumption allows us to write
the following equations:
\beq
\frac{{\widetilde \phi(x_{M+1},t_n)}}{{\widetilde \phi(x_{M},t_n)}}=\frac{{\widetilde \phi(x_{M},t_n)}}{{\widetilde \phi(x_{M-1},t_n)}}
=\exp(ik_x\Delta x)
\eeq
From the second equality one obtains  $k_x$ and then from the first equality it results 
\beq
{\widetilde \phi(x_{M+1},t_{n})}={\widetilde \phi(x_{M},t_{n})}\exp(ik_x\Delta x)
\eeq
We suppose the same relation valid for the next time step, so that 
\beq
{\widetilde \phi(x_{M+1},t_{n+1})}={\widetilde \phi(x_{M},t_{n+1})}\exp(ik_x\Delta x)
\eeq

Formula (\ref{fluxr}) shows that an outgoing or ingoing flux 
developes across the boundary depending on the sign of 
${\rm Real}\left (\hbar k_x \right ) +eZ_{\rm eff}A(t)$. 
Due to the action of the time-dependent field, the real part of $k_x$ is 
no longer constrained to be always positive, as happens for $\alpha$-decay 
in the absence of an external perturbation \cite{mis00}. For this last process
the overall change in energy from the right boundary is always 
negative and thus the wave function flows out of the grid region. In the present  case, back-flow of 
the wave-function inside the numerical domain is expected to occur due to the reversal of the electric field polarity. 
In practice we can compute the outgoing flux and add it to the norm if a part of the wave function goes out of the domain.
We therefore have full control on the  norm used in the calculation of tunneling probability and decay rate. 
Employing the TBC procedure one can obtain values of the physical
quantities with a much smaller extension of the spatial grid than is necessary
without the TBC procedure. 
  
To solve TDSE an initial wavefunction has to be constructed at $t=0$. 
Like in our previous paper \cite{mis00} we use a recipe \cite{jackbrown77}
that provides this initial wavefunction, $\phi_0(x)$, as a bound state of the 
stationary Schr\"odinger equation in the modified static potential (the potential 
$V(x)$ having a constant value $V(\pm x_{\rm mod})>E_{\alpha}$ for a distance $|x|>|x_{\rm mod}|$
beyond the top of the barrier) \cite{jackbrown77}. Accordingly, the intial wave-function is
found by considering the stationary Schr\"odinger equation without any perturbation (${\cal E}_0=0$)  
\beq
\frac{d^2 \phi}{d x^2}+\frac{2\mu}{\hbar^2}\left [E_{\alpha}-V_{\rm mod}(r) \right ]\phi = 0
\label{eigvaleq}
\eeq  
where $V_{\rm mod}(r)$ is the modified potential. 
The above equation is solved by discretising the space coordinate with the same difference formula for the second
derivative and thus transforming it to a matrix eigenvalue problem. The depth $V_0$ of the WS-potential is varried until
one of the eigenvalues is matching the experimental decay energy $E_\alp$.

{red}{At early times of the quantum tunneling process the decay rate deviates strongly from the 
exponential law. This is a common feature of many unstable quantum systems, not only of $\alp$ emitters 
\cite{fonda78,wilk97}.
As shown above, the wave function at $t=0$ is prepared as a bound eigenstate of the modified potential.
It is therefore a wave-packet with a sharp energy in contrast to the wave-packet at $t>0$, which
possess an energy width $\Delta E\neq 0$ due to the interaction responsible for the decay, 
i.e. the unmodified potential. After the "injection" at $t=0$ of the initial wave-function in the real Hamiltonian, there will be a transient period necessary for the wave-packet to adapt to its new environment.
In this transient time, $\tau_{\tr}$, which is much less than the decay time, the decay is expected to be non-exponential. This behavior was numerically proved in our earlier work \cite{mis00} where 
we concluded that if the decay energy decrease, the time necessary for the decay rate to reach its 
asymptotic value (exponential behavior regime) increase. For decays near threshold there is no exponential decay regime.
After the transient regime it is natural to expect that the probability density inside the numerical
grid (\ref{pinterior}) obeys the radioactive decay law
\beq
P_{\rm int}(t)=P_{\rm int}(0)e^{-\lam t}
\eeq   
Taking the time derivative of the above formula and inserting eq.(\ref{balans}) 
we obtain the decay rate in terms of the quantum flux balance between the interior 
(inside the numerical grid) and the exterior (outside the numerical grid) domains    
\beq
\lambda(t)=-\frac{1}{P_{\rm int}(t)}({J({x_{\rm left}},t)-J({x_{\rm right}},t)})
\label{decrat}
\eeq
Note that in all numerical experiments performed in the next section, the laser wave period is 
larger than the transition time.}

A quantity that provides a measure of the tunneling speed, frequently used in atomic physics, is the 
``ionization`` probability \cite{chou12}
\beq
P_{\rm ion}=1-\left | \int_{x_{\rm left}}^{x_{\rm right}}{\widetilde \phi^*}(x,t) \phi_0(x)\right |^2
\eeq
It provides the probability that the time-dependent metastable state of the $\alpha$ particle, ${\widetilde \phi}(x,t)$, 
departures from the initial bound state $\phi_0(x)$. At this point we remind the reader that
the solution of the corresponding stationary Schr\"odinger equation with purely outgoing boundary conditions (Gamow state), 
$\psi(x,t)$ has a complex eigenvalue, $E_{\alpha}-i\hbar\lam/2$ and decays exponentially in time \cite{brand89}
\beq
L(t)=\left | \int_{x_{\rm left}}^{x_{\rm right}}{\psi^*}(x,t) \psi(x,0)\right |^2=\exp[-\lam t]
\eeq
The decay rate $\lam$ from the above formula should be interpreted as the asymptotic value of (\ref{decrat}), i.e. 
$\lam=\lam(\infty)$

\section{Numerical experiments}
 
Let us review first the magnitudes of the laser parameters employed in this study.
We proceed in close analogy to atomic physics \cite{muls10} with the difference that 
we use nuclear adapted units (for energy, time, size) instead of atomic units.  
 
The relation between the electric field strength ${\cal E}_0$ and the laser intensity 
$I_0=\frac{1}{2}c\epsilon_0 {\cal E}_0^2$ can be recasted for practical purposes 
as ${\cal E}_0$[V/cm]$=27.44 \left \{ I_0 [{\rm W}/{\rm cm}^2]\right \}^{1/2}$ \cite{muls10}. It shows  
that at the maximum intensity foreseen at ELI \cite{eli09}, i.e. $I_0\approx 10^{25}$W/cm$^2$, 
the electric field is ${\cal E}_0\approx8.7\times10^{13}$V/cm or, in nuclear units, 
$e{\cal E}_0\approx$ 8.7$\times10^{-6}$MeV/fm.
In this paper we explore a wide range of laser intensities above this value.
We find convenient to express 
the frequency in nuclear units as $\hbar\omega=1240 [{\rm MeV\cdot fm}]/\lam[{\rm fm}]$. 
Thus for a X-ray laser with wavelength of 
$\lam=1.24$ \AA, the photon energy is $\hbar\omega$=10 keV and the period $T=2\pi/\omega=4.13\cdot10^{-19}s=0.413$ attoseconds(as).
It means that the present study employs a laser with wavelengths in the range of XFEL, and as we shall see below 
even smaller. A radiation produced by a Ti:saphire laser, correspond to a period of $T$=2.66 fs, i.e. of the same
order of magnitude as the pulse duration, but also close to the maximum time limit of the TDSE solution. It means
that with this type of laser the evolution of the decaying system can be followed only during a single period of the light signal.    
Our intensities are beyond the highest value expected at ELI and this is because we intend to magnify the effect of a 
superstrong laser pulse on the radioactive nuclear system and thus grasp its salient features.
In other words the present study is carried out assuming parameters for a fictitious laser facility, which is however 
not far beyond the technical possibilities foreseen in the near future. For that one would neccesitate pulses as short 
as 0.1 fs of X-rays of wavelength 1 \AA~focused into a  1$\times$1 nm$^2$ spot. For 10$^{12}$ photons (10 times 
larger than at SwissFEL), an energy release of $10^{-2}$ J, a power peak of 0.1 PW and an intensity of 
$I_0=10^{28}$ W/cm$^2$ could be attained.

As we already mentioned, the response of the $\alpha$-daughter system to the impinging laser field 
is dominated by the electric-dipole term : -$Z_{\rm eff} e \bi{E(t)}\cdot\bi{r}$. To apprehend the 
importance of this coupling, as compared to excitations of the $\alpha$-daughter system (such as excitations
from the ground state $E_{\alp}$ to higher-lying states in the continuum), we introduce in full analogy to the  
atom-laser interaction case \cite{muls10}, the maximum value of the {\em ponderomotive energy}, 
$U_p=(Z_{\rm eff} e {\cal E}_0)^2/(4\mu\omega^2)$, which corresponds to the averaged kinetic energy gained by a 
free (i.e. $V(r)$=0) particle of mass $\mu$ and effective charge $Z_{\rm eff}$ once it is set in forced oscillations 
by the laser field. For an intensity of $10^{29}$W/cm$^2$
and $\lam=1.24$ \AA~  applied to the $\alpha$-radioactive
nucleus $^{106}$Te, we get 
$U_p\approx2.9\cdot10^{-5}$ MeV. 
When $U_p\approx \mu c^2$, relativistic effects are expected to be important. 
The square of the dimensionless parameter  
$\eta={v}/c=Z_{\rm eff} e{\cal E}_0/(\mu\omega c)=\sqrt{4U_p/\mu c^2}$, 
that represents the ratio of the maximum quiver velocity ${v}$ of the laser irradiated decaying system to 
the velocity of light, is another parameter that indicates the onset of relativistic effects, when $\eta^2\geq$1.
{ For the above selection of XFEL laser parameters the non-relativistic regime
is still effective since $\eta\approx 1.8\cdot 10^{-4}$. For the Ti:saphire laser however, the 
wavelength is almost four orders of magnitude larger and for the same intensity the ponderomotive potential 
$U_p\approx 1200$ MeV approaches $\mu c^2$ and consequently the overunitary value of $\eta=1.16$ signalizes 
the come into play of relativistic effects.} 

We specify the parameters of the nuclear potential according to the $\alpha$-nucleus optical potential compilation from 
\cite{ripl-alfa}. For $^{106}$Te, the parameters of the WS potential are : $V_0$=-137.7 MeV, $a$=0.76 fm and $r_0$=1.235 fm
and are consistent with the energy $E_\alpha$=4.15 MeV if one solves the stationary Schr\"odinger equation (\ref{eigvaleq})
with the potential modified at $|x_{\rm mod}|$=25 fm. The spatial border is specified by $x_{\rm right,left}=\pm$192 fm.

\begin{figure}
\begin{center}
\epsfbox{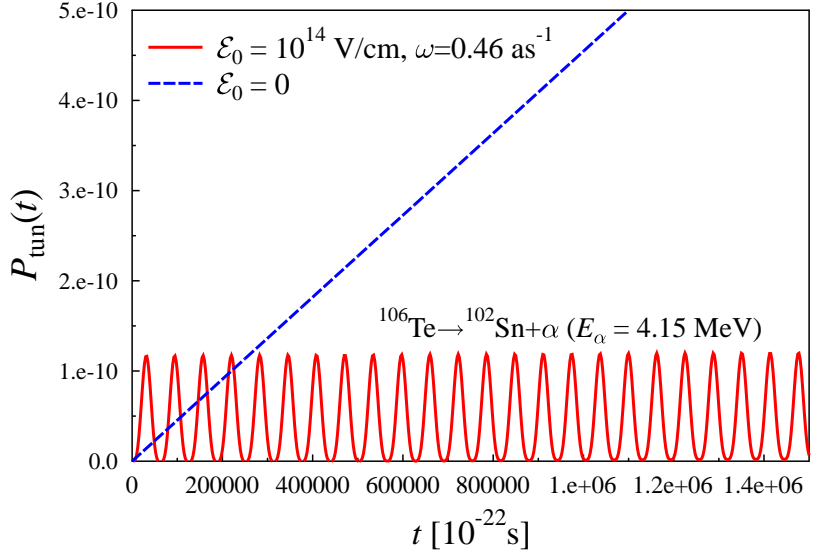}
\end{center}
\caption{(Color online) Tunneling probabilities of the $\alpha$ particle from $^{106}$Te for a continuous pulse of frequency 
$\omega=$0.46 as$^{-1}$ and amplitude 
${\cal E}_0=10^{14}$V/cm or $e{\cal E}_0= 10$ eV/fm 
and when the laser field is turned off (blue dashed curve)}
\label{fig1}
\end{figure}

As an envelope function we use a sequence of square pulses
\beq
{F}(t)=\sum_{p=0}^{N}(-)^p\theta(t-\tau_p)
\eeq
such that the laser pulses act in the time intervals [$\tau_0=0,\tau_1$],[$\tau_2,\tau_3$], etc. We recover the cw case when 
$\tau_0\longrightarrow \tau_{p}$, where $\tau_{p}$ is the duration of a single long pulse with constant amplitude. 
In the general case we consider short pulses with constant amplitude 
of length equal to an integer number of half-cycles and separated by intervals of similar length, i.e. if at the time $\tau_{2i}$
the field is again turned on and then at $\tau_{2i+1}$ is again turned off then the duration of the $(i+1)$th short pulse is 
\beq
\tau_{i+1}-\tau_{i}=n_{i+1}\frac{T}{2}
\eeq   
where $n_{i+1}$ is the number of half-cycles .  

Let us study first the effect of the radiation for a continuous sinusoidal pulse on the dinuclear system.  
In Fig.\ref{fig1} we compare the tunneling probability in the unperturbed case and in the case when properties 
of the laser impinging on the decaying system are given by an electric field strength of ${\cal E}_0=10^{14}$V/cm 
($e{\cal E}_0=10$ eV/fm), which corresponds to the intensity $I_0=1.33\cdot 10^{25}$ W/cm$^2$, just slightly above the ELI maximum value
and frequency $\omega\approx 4.6 10^{17}{\rm s}^{-1}$
($\hbar\omega\approx $0.3 keV) which translated in wavelength is $\lambda\approx$41.3 \AA. 

It can be inferred at first glance that even at a rather low field intensity, compared to the barrier ($e{\cal E}_0\ll V_{\rm barrier}$),
the escape chances of the $\alpha$ particle are not only diminishing but on a long term the nuclear $\alpha$-daughter dipole 
is oscillating around an average value that increases with a very small slope. 

At this point we recall the reader that a phenomenon which resemble the one studied in this article, is the well-known  
ionization of an electron moving in a static Coulomb potential and an intense laser field \cite{gavr92,muls10}. 
The effect of the varying electric field is to lower over half of cycle the barrier seen by the electron and thus enhance 
the ionization probability by escape  over the {\em downhill} barrier.  
Thus, in analogy to the case of atomic ionization we can speak of a
stabilization of the decaying system due to the action of the radiation field. 

\begin{figure}
\begin{center}
\epsfbox{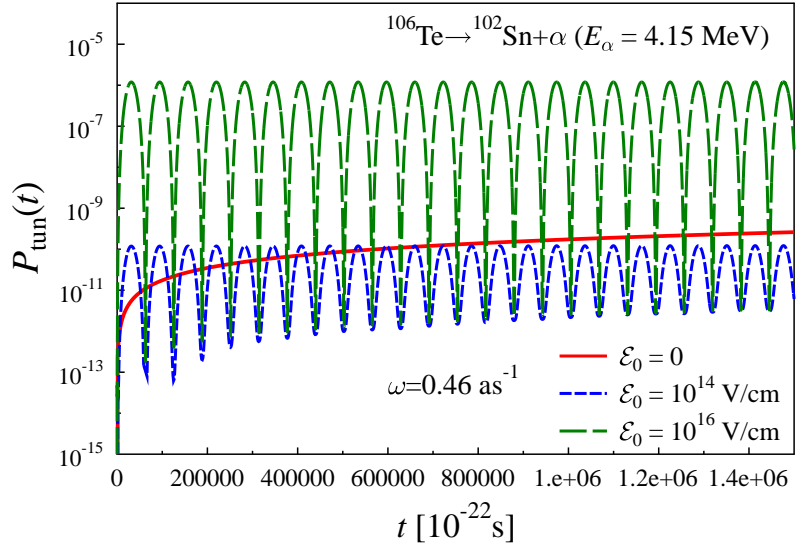}
\end{center}
\caption{(Color online) Tunneling probabilities in the logarithmic scale for two different cw 
field amplitudes ${\cal E}_0=10^{14}$V/cm and $10^{16}$V/cm 
and frequency $\omega=$0.46 as$^{-1}$ are compared to the tunneling probability without laser field.}
\label{fig2}
\end{figure}

The effect of increasing ${\cal E}_0$ for a fixed frequency can be visualized in Fig.\ref{fig2}.  
To that end we compare the tunneling probabilities for the field-free case to the case when 
a cw laser with the same frequency as in Fig.\ref{fig1} acts on the decaying system with various
intensities. We infer from this figure that the higher the intensity the higher
is the oscillation amplitude of $P_{\rm tun}(t)$. From classical point of view this effect is related to the oscillation of 
the alpha particle in the varying electric field. In this picture recollision happens when $P_{\rm tun}(t)$ approaches its 
minima.

\begin{figure}
\begin{center}
\epsfxsize=41pc
\epsfbox{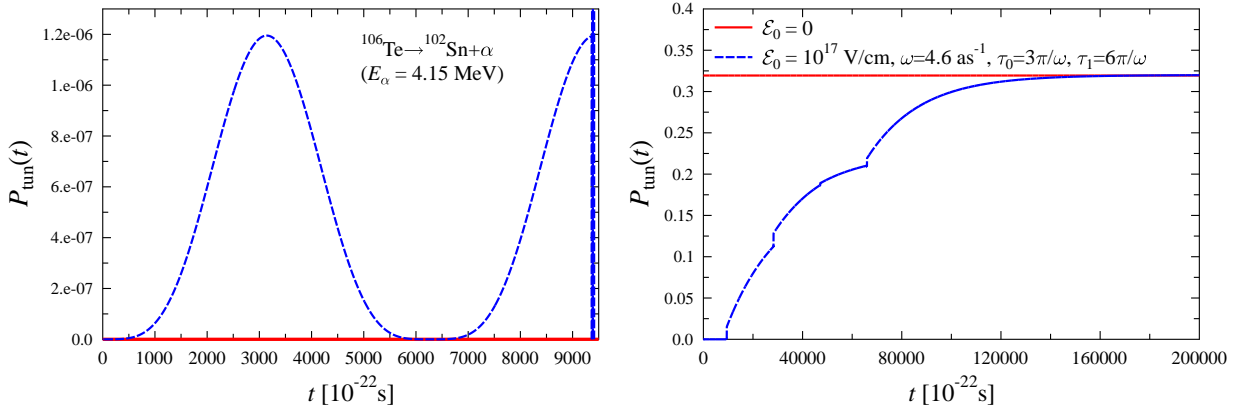}
\end{center}
\caption{(Color online) Tunneling probabilities of the $\alpha$ particle from $^{106}$Te for  ${\cal E}_0$=10 keV/fm and $\omega$=4.6 as$^{-1}$
for a sequence of 4 pulses of duration 3$\pi/\omega$ (separated by three breaks of the same duration) is compared to the field-free case.
 On the left panel the time axis runs up to 950 zs. Note that after a complete cycle (2$\pi/\omega$) the tunneling probabilities 
 are similar. On the right panel the curves are represented in the time interval extended to 20 as and the field-free $P_{\rm tun}(t)$ is multiplied
by a factor of 850.}
\label{fig4}
\end{figure}

Let us next consider a laser field perturbing the decaying  system as a sequence of abruptly turned-on short pulses consisting of an 
odd number of half-cycles. 
In Fig.\ref{fig4} we compare the tunneling probability for a laser signal formed  of 4 pulses, 
each of length 3$\pi/\omega$ and for the field-free case. We use this time a huge  
field of amplitude ${\cal E}_0=10^{17}$V/cm ($e{\cal E}_0=10$ keV/fm) and a smaller wavelength of $\lam=4.13$\AA~for the sole purpose 
of giving the reader a clear picture of the tunneling probability enhancement. 
As can be seen on the right panel of this figure, which represents the large time scale behavior, the factor gained by 
$P_{\rm tun}(t)$ when applying an odd number of 
half-cycles is $\sim$850 after $\sim$150 as.

\begin{figure}
\begin{center}
\epsfxsize=27pc
\epsfbox{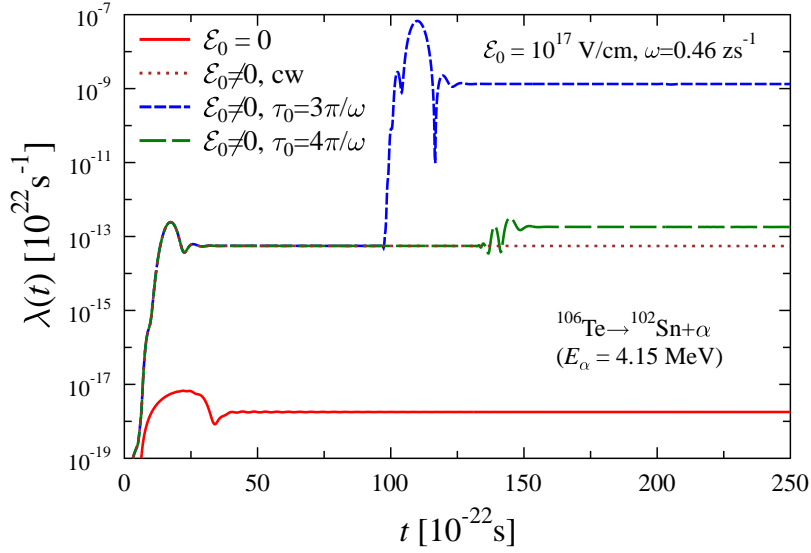}
\end{center}
\caption{(Color online) Decay rates of the $\alpha$ particle from $^{106}$Te for 4 cases : field-free case (full curve), continuous wave (dotted ),  
short pulse of 3 half-cycles duration (short dashes), short pulse of 4 half-cycles duration (long dashes). The values of the radiation field and 
frequency are  ${\cal E}_0$=10 keV/fm and $\omega$=0.46 zs$^{-1}$. Note that the repetition of the pulse is made after a time equal to the corresponding
pulse duration.}
\label{fig5}
\end{figure}

In Fig.\ref{fig5} we tried to represent in a suggestive manner how the decay rates (\ref{decrat}) are step-wise increased  when the electric field 
is  turned off. This behavior is caused by the box-like shape of the laser signal envelope. For a gaussian or $\sin^2$ envelope we should expect
a smooth increase of $\lambda(t)$. This could be the theme of a future study. 
We compare here the pulse with an odd number of half-cycles with the pulse containing an even number of half-cycles. The jump of $\lambda$ in the
second case is much less pronounced compared to the first case when the laser field is turned off. 
Since the half-live $T_{1/2}$ is inverse proportional to the asymptotic 
value of  $\lambda$, Fig.\ref{fig5} tells us that in the most favorable case, e.g. when a very short pulse with an odd number (3) of half-cycles is 
interacting with the decaying system, $T_{1/2}$ decrease by $\sim$ 9 orders of magnitude! Otherwise stated, the decay can be speeded up in the order 
of femtoseconds instead of microseconds for the $\alpha$-radioactive nucleus $^{106}$Te.     

\begin{figure}
\begin{center}
\epsfxsize=40pc
\epsfbox{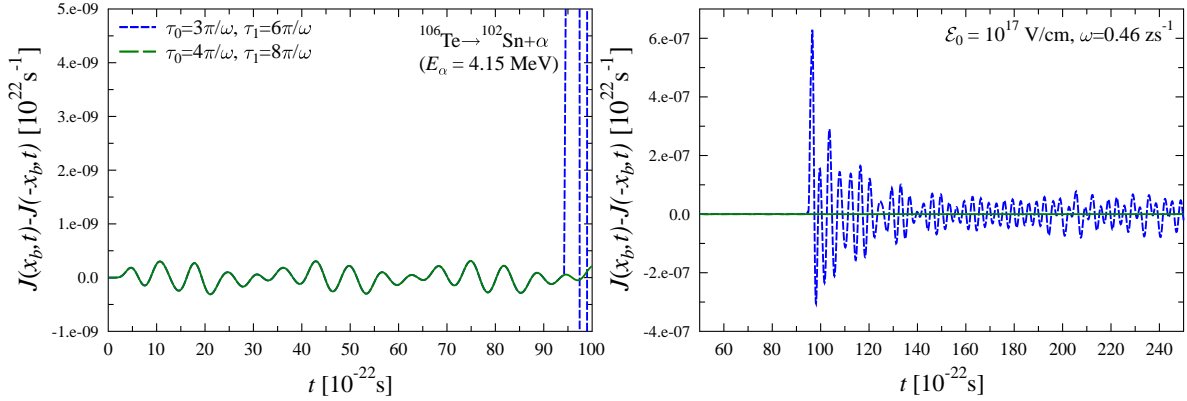}
\end{center}
\caption{(Color online) Total flux across the nuclear domain in the case of a  
short pulse of 3 half-cycles duration (short dashes), and a second one of 4 half-cycles duration (long dashes). The values of the radiation field and 
frequency are  ${\cal E}_0$=10 keV/fm and $\omega$=0.46 zs$^{-1}$. Note that the repetition of the pulse is made after a time equal to the 
corresponding pulse duration.}
\label{fig6}
\end{figure}

{
To understand the jump in the decay rates we return to formula (\ref{decrat}).  
The denominator of the fraction in the r.h.s. of this equation, which represents the probability to find the 
particle inside the numerical domain at a given time, is certainly experiencing an insignificant 
variation at early times, especially for a large integration domain like the one used by us.
Remember that the wave-packet at time $t$=0 is confined in the nuclear domain.    
Therefore when the pulse with an odd number of half-cycles is turned off, the decay rate owe its 
abrupt variation  solely to the disbalance in the flux. The jump is even more pronounced if we evaluate 
the total flux leaving or returning inside the nuclear surface, i.e. we take as reference the 
left and right Coulomb barrier positions. This disbalance between the flux at the left border 
and the one at the right border is caused by the interuption of the pulse at the half of a period. 
Would have been the pulse continued up to the completion of the period, then the contributions from both 
ends would almost compensate each other. As can be inferred from Fig.\ref{fig6}, once the alternating 
electric field is turned off, the quantum flux affected by the pulse with an odd number of half-cycles is 
oscillating with a much larger amplitude compared to the pulse with an even number of half-cycles.}

\section{Conclusions and perspectives}

In this paper we proposed a numerical algorithm based on the CN method
to solve the TDSE and therefore to examine the dynamics of the
$\alpha$-decay process under the influence of an ultra-intense monochromatic
laser field.
Our primary goal was to establish the characteristics of the laser pulse that
entails a major modification  of the tunneling probabilities, decay rates and thus of half-lives.
The most important result of our study was that short pulses containing an odd number of half-cycles instead
of an even number are massively affecting this type of nuclear radioactivity. A repeated application of
such pulses leads to a faster decay of an $\alpha$-radioactive nucleus. We proved that in a time below an
attosecond it was possible to increase the decay rates by several order of magnitudes.    
To substantiate this effect we used in our theoretical study ultra-intense laser fields that are awaiting 
to be produced by the next generations of laser facilities.  
Laser control of nuclear decay processes is a new facette of the emerging field of direct laser-nucleus reactions.
A better knowledge of the mechanism governing this type of phenomena could also find applications in the domain of
radioactive waste disposal by using high-power lasers.

Another effect that we cannot exclude apriori is  that the free cluster-daughter dipole, driven by the laser field, 
is prone to emit electromagnetic radiation. We already investigated more than a decade ago the bremsstrahlung in 
standard $\alpha$-decay and concluded that the contribution coming from the tunneling could be present at most
in the region of hard photons \cite{mis00}. In a recent publication \cite{corso03} it was conjectured that a charge moving 
in the vicinity of a stationary scattering center and subjected to a strong laser field, emits  electromagnetic
radiation close to the scatterer in a well-defined time interval.
In this way it is possible in principle to determine the $\alpha$-particle position by observing the radiation 
that it is emitting.     
  
Our contribution also calls the attention on the possibility to control spontaneous radioactive decays with 
super-strong electromagnetic fields

\ack
Thanks are indebted to dipl.eng. C. Matei from NILPRP Bucharest-Magurele 
for carefully reading the manuscript and her valuable observations. 
We acknowledge the financial support from UEFISCDI Romania under the programme 
PN-II contract no. 116/05.10.2011.

\end{document}